\documentstyle[12pt,epsf]{article}
\begin{document}
\begin{center}
{\Large \bf Neutrino masses and $Z^\prime$ physics}

\vspace*{4mm}
V.V.Kiselev\\
Institute for High Energy Physics\\
142284, Protvino, Moscow Region, Russia.\\
E-mail: kiselev@mx.ihep.su
\end{center}

\begin{abstract}
The characteristic scale of neutral current, provided by an extension 
of Standard Model with a local group over the right-handed fermions,
determines the smallness of neutrino masses of Dirac kind. The experimental
observation of neutrino oscillations imposes the stringent limit on
the $Z^\prime$ physics appearance at low energies.
\end{abstract}

\section{Introduction}
The recent clear observation of neutrino oscillations \cite{SK} 
indicates the non-zero masses for the neutrinos \cite{inter}, 
which are generally considered to be extremely small due to the 
following mechanisms. The common form of neutrino mass matrix is 
expressed through three terms:
\begin{enumerate}
\item 
the Majorana mass $m_T$ for the left-handed neutrino, that appears as 
the triplet weak isospin contribution,
\item
the Dirac mass term $m_D$, involving the interaction with the sterile 
right-handed neutrino $\nu_R$,
\item
the Majorana mass $m_S$ for the singlet $\nu_R$ over the weak interaction.
\end{enumerate}
Then the Lagrangian part, determining the neutrino mass, is equal to
\begin{equation}
L_m = (\bar \nu_R,\; \bar \nu_L) 
\left(\begin{array}{cc} m_S & m_D \\ m_D & m_T \end{array}\right)
(1+C) \left({\nu_R \atop \nu_L}\right) +{\rm h.c.},
\end{equation} 
where $C$ denotes the charge conjugation. The following possibilities 
are generally discussed.

The first is the absence of sterile $\nu_R$, that means there is 
the only contribution due to the Majorana term $m_T$. If it is 
caused by the vacuum expectation $v$ of standard higgs, being the isodoublet,
then the triplet mass is the square of $v$. Hence, $m_T\sim v^2/M$, where
$M$ appears from an extension of SM, and it is of large scale.

Second, the singlet contribution is determined by the physics beyond the SM,
so that $m_S \gg m_D \gg m_T$, where $m_D$ is usually taken in the range,
corresponding to the mass scale for the charged fermions, $m_D \sim v$. Then
the see-saw mechanism \cite{seesaw} leads to two spices of neutrinos with 
the small and large masses, $m_1 \sim m_D^2/m_S$ and $m_2 \sim m_S$, 
correspondingly.

These scenarios (just beyond a super-physics) generally exhaust 
the natural explanations for the 
smallness of neutrino masses. Anyway, the experimental data put 
$m_\nu > 0.01$ eV, which means $M\sim m_S \sim 10^{15}$ GeV.

Note, that we do not know how the small Dirac masses can be naturally
explained with no involvement of large sterile Majorana mass.

In the present paper we offer the scheme, wherein the neutrinos have
zero masses in the Standard Model and acquire the smallness after 
an extension of SM to include the right-handed neutral currents. We suggest
some non-trivial vacuum correlators, which determine the mass scales in
connection to the gauge charges of fermions. Thus, it happens that the
large scale $M$ is related to the mass of $Z^\prime$.

\section{Mass generation}

The second order contribution of SM to the effective action contains 
the neutral current term of the form
\begin{eqnarray}
  S_{2m} &=& \int dx dy\;  \frac{e}{\cos \theta \sin \theta}
(T_3-Q \sin^2 \theta)[\bar L_L(x) Z_\mu(x) \gamma^\mu L_L(x)] 
\cdot \nonumber \\&& ~~~~~~
Q {e}\;{\tan \theta} [\bar L_R(y)Z_\nu(y) \gamma^\nu L_R(y)],
  \label{eq:2}
\end{eqnarray}
where we introduce the notations $L_L$ for the left-handed doublets and
$L_R$ for the right-handed singlets. Further, suggest the non-trivial 
vacuum correlators with the characteristic distance $r\sim 1/v$
\begin{eqnarray}
  \langle 0| Z_\mu(x) \gamma^\mu L_L(x)\; 
\bar L_R(y) Z_\nu(y) \gamma^\nu |0\rangle  &=&
\frac{\delta(x-y)}{v^4} \langle 0| Z_\mu(x) \gamma^\mu L_L(x)\; 
\bar L_R(x)  Z_\nu(x) \gamma^\nu |0\rangle \nonumber \\ & \sim &
\delta(x-y)\; v ,
  \label{eq:3}
\end{eqnarray}
where we suppose that the scales of expectations for $ZZ$ and 
$L_L \bar L_R$ equal $v^2$ and $v^3$, respectively.
Therefore, the fermion masses of Dirac kind are determined by the action
\begin{equation}
  \label{eq:3a}
  S_{fm} \sim \int dx\; \bar L_L(x) L_R(x)\cdot v\cdot
 \frac{e^2}{\cos^2 \theta}(T_3-Q \sin^2 \theta) Q+{\rm h.c.}
\end{equation}

From (\ref{eq:3a}) we deduce that the coupling of vacuum expectations,
causing the Dirac masses, is determined by the charges of fermions, so that,
say, for the neutrino the electric charge equal to zero results in the
massless, which, thus, looks quite natural in the SM with the suggested 
mechanism for the mass generation. 

Sure, we could introduce the local source, i.e. the Higgs field, for 
the vacuum expectation considered in the model above and make the 
Legendre transformation to substitute the field for the condensates. Then, we
believe, the action would take the most usual form\footnote{In this way we 
have, say, to consider the initial bi-local bare $J$-action 
$$S_{ib}=\int dx dy J(x,y)\; [\bar L_R(x)\; Z_\mu(x) \gamma^\mu 
Z_\nu(y) \gamma^\nu \; L_L(y)] - \int dx \phi(x) J(x,x)+{\rm h.c.}$$
To the bare 
order the equation of motion for the bi-local field results in the 
straightforward substitution of local field $\phi$, as it stands in the 
above consideration for the correlators, developing the vacuum expectation 
values. After the analysis of divergences in the $J$-dependent 
Green functions, the corresponding counter terms must be added to the action. 
Then the $J$-source can be integrated out, that believes to result in the 
$\phi$-higgs action, containing the couplings to fermions as well as the
suitable potential to develop the spontaneous breaking of symmetry.}.

\subsection{Neutrino masses from right-handed extension}

Suppose that there is the additional $SU(2)_R$ local symmetry spontaneously
broken at a scale $v_R$ (for the sake of explicitness we put $v=v_L$ and
$v_R^2 = b\; v_L^2 \gg v_L^2$). To  minimize possible virtual corrections at 
low energies precisely studied up to the LEP measurements and to reproduce
the mass relations between the standard gauge bosons, we have to 
introduce the additional Higgs field, being the doublet over $SU(2)_R$, which
possesses zero charges over $U(1) \otimes SU(2)_L$ of the SM. The important
challenge is that the standard higgs has to be extended to the field, 
belonging to the ($\frac{1}{2},\frac{1}{2}$) representation of 
$SU(2)_L\otimes SU(2)_R$. The reason is the desirable renormalizability of
the field theory with the spontaneously broken symmetry. Indeed, the fermion 
mass term $\bar \psi_R \psi_L$ certainly is of ($\frac{1}{2},\frac{1}{2}$), 
so that the corresponding higgs developing the {\sc vev} must have the same
quantum numbers over the local group\footnote{For the subject under 
consideration it is not so significant that the extension is $SU(2)_R$. It 
can be, say, $U(1)_R$. In the current discussion we can suppose that the 
right-handed doublets are $\left(e \atop d\right)$ and $\left(\nu
\atop u\right)$. Then the charged $SU(2)_R$-like gauge bosons are the vector 
leptoquarks, that can be the reason for the very different physics in the
left- and right-handed sectors.}.

Next, the non-zero {\sc vev}s of neutral Higgs fields result in the massless
of the photon and lead to the massive neutral currents. The mass matrix for
the local gauge fields $B$, $Z_L$ and $Z_R$ is determined by the form
\begin{equation}
  \label{eq:4}
  M^2_{gauge} =\frac{1}{4}{v_L^2} \left(
    \begin{array}{ccc}
  g^2 & - g g_L & g g_R \\
  -g g_L & g_L^2 & - g_L g_R \\ 
  g g_R & -g_L g_R & g_R^2 (1+b)
    \end{array}
\right),
\end{equation}
where $g$, $g_L$ and $g_R$ denote the gauge fields couplings for the
$U(1)\otimes SU(2)_L\otimes SU(2)_R$ group.

Then the eigenvalues of the mass matrix are equal to
\begin{eqnarray}
  m_A^2 & = & 0, \nonumber \\
  m_Z^2 & = & \frac{1}{8}{v_L^2}\left(\sum_i g_i^2+b g_R^2 -
\sqrt{(\sum_i g_i^2)^2 - 2b g_R^2(g^2+g_L^2-g_R^2)+b^2 g_R^4}\right), \\
  m_{Z^\prime}^2 & = & \frac{1}{8}{v_L^2}\left(\sum_i g_i^2+b g_R^2 +
\sqrt{(\sum_i g_i^2)^2 - 2b g_R^2(g^2+g_L^2-g_R^2)+b^2 g_R^4} \right),
\nonumber
\end{eqnarray}
which in the limit of infinitely large $b$ tend to the following relations:
\begin{eqnarray}
  m_A^2 & = & 0, \nonumber \\
  m_Z^2 & \approx  & \frac{1}{4} v_L^2 (g^2+g_L^2) (1-\frac{1}{b})
\approx \frac{1}{4} v_L^2 (g^2+g_L^2), \\
  m_{Z^\prime}^2 & \approx  & \frac{1}{4} [(v_R^2+v_L^2) g_R^2+v_L^2 
(g^2+g_L^2)\frac{1}{b}] \approx \frac{1}{4} v_R^2 g_R^2.
\nonumber
\end{eqnarray}
Furthermore, it is quite evident to derive that the masses of charged
gauged bosons are given by
\begin{eqnarray}
  m_{W_L}^2 & = & \frac{1}{4} v_L^2 g_L^2, \nonumber \\
  m_{W_R}^2 & = & \frac{1}{4} [v_R^2 g_R^2+v_L^2 g_L^2].
\nonumber
\end{eqnarray}
Thus, we see that up to small corrections the $Z$ boson mass reproduces the
value of standard boson as it is connected to the $W$ mass.

The matrix $S$, transforming the gauge fields to the mass eigen-states, 
has the form
\begin{equation}
  \label{eq:6}
  \left(
    \begin{array}{c}
A \\ Z\\ Z^\prime
    \end{array}\right) = \left(
    \begin{array}{ccc}
\cos \theta & \sin \theta & 0 \\
-\sin \theta &\cos \theta & \frac{1}{\beta} \\
\frac{1}{\beta} \sin \theta  & -\frac{1}{\beta} \cos \theta  & 1
    \end{array} \right) \left(
    \begin{array}{c}
B \\ Z_L \\ Z_R 
    \end{array}\right),
\end{equation}
with the accuracy up to $O(\frac{1}{\beta^2})$, 
where $\beta=b {g_R}/\sqrt{g^2+g_L^2}$,
and $\theta$ is the standard angle by Weinberg. In the approximation under
consideration we can see that $S$ has the orthogonal form, and 
the transposition results in the inverse matrix, 
\begin{equation}
  \label{eq:6b}
\left(
    \begin{array}{c}
B \\ Z_L \\ Z_R 
    \end{array}\right) = \left(
    \begin{array}{ccc}
\cos \theta & -\sin \theta & \frac{1}{\beta} \sin \theta \\
\sin \theta &\cos \theta & -\frac{1}{\beta} \cos \theta \\
0  & \frac{1}{\beta}  & 1
    \end{array} \right) \left(\begin{array}{c}
 A \\ Z\\ Z^\prime
    \end{array}\right) ,
\end{equation}
so that the admixture of $Z^\prime$ in $Z_L$ is determined by the
ratio of $v_L^2/v_R^2$.

The vertices of massive eigen-states are determined by the following 
relations:
\begin{enumerate}
\item 
The photon couples to the electric charge 
$$
Q = \frac{Y^L}{2}+T^L_3=\frac{Y^R}{2}.
$$
\item 
The left-handed fermions have the standard couplings to $Z$
$$
\frac{e}{\cos\theta\sin\theta}(T_3^L-Q \sin^2 \theta).
$$
\item 
The right-handed fermions acquire the correction to the charge with $Z$
$$
-Q e \tan \theta + T^R_3 \frac{g_R}{\beta}.
$$
\item 
The $Z^\prime$ vertex to the left-handed fermions is proportional to that of 
$Z$ one, so that
$$
-\frac{e}{\beta\cos\theta\sin\theta}(T_3^L-Q \sin^2 \theta),
$$
and the suppression is due to the smallness of $v_L^2/v_R^2$.
\item 
$Z^\prime$ has the charge
$$
T^R_3 g_R +Q e\frac{\tan \theta}{\beta},
$$
to the right-handed fermions.
\end{enumerate}
So, some anomalous couplings are introduced due to the $Z^\prime$ physics.

Now suggesting the non-trivial vacuum correlations at the distances of
$r\sim 1/v_R$ we find that the neutrinos acquire the non-zero masses due to
the admixture of $Z^\prime$ in $Z_L$ and its dominance in $Z_R$. So, the action
\begin{eqnarray}
  S^{\prime}_{2m} &=& - \int dx dy\;  \frac{e}{\beta\cos \theta \sin \theta}
(T^L_3-Q \sin^2 \theta)[\bar L_L(x) Z^{\prime}_\mu(x) \gamma^\mu L_L(x)] 
\cdot \nonumber \\&& ~~~~~~
\left(T^R_3 g_R+Q {e}\frac{\tan \theta}{\beta}\right) 
[\bar L_R(y)Z^{\prime}_\nu(y) \gamma^\nu L_R(y)],
  \label{eq:7}
\end{eqnarray}
transforms to
\begin{equation}
  \label{eq:7a}
  S^{\prime}_{fm} \sim \int dx\; \bar L_L(x) L_R(x)\cdot \frac{v_L^2}{v_R}
\cdot  \frac{e g_R}{\beta\cos \theta\sin\theta}
(T^L_3-Q \sin^2 \theta) T^R_3 +{\rm h.c.}
\end{equation}
if
\begin{eqnarray}
  \langle 0| Z^{\prime}_\mu(x) \gamma^\mu L_L(x)\; 
\bar L_R(y) Z^{\prime}_\nu(y) \gamma^\nu |0\rangle  &=&
\frac{\delta(x-y)}{v_R^4} \langle 0| Z^{\prime}_\mu(x) \gamma^\mu L_L(x)\; 
\bar L_R(x)  Z^{\prime}_\nu(x) \gamma^\nu |0\rangle \nonumber \\ & \sim &
\delta(x-y)\; v_R ,
  \label{eq:8}
\end{eqnarray}
where we suppose that the scales of expectations in these correlations
for $Z^{\prime}Z^{\prime}$ and $L_L \bar L_R$ equal $v_R^2$ and $v_R^3$, 
respectively.

Note, that, first, other correlations result in less contributions 
to the masses, as those are suppressed by powers of $v_L/v_R$.
Second, the corrections to the masses of electrically charged fermions seem
to be suppressed in the same manner.

Thus, we see that due to the extension of model to the right-handed local
group the neutrinos have the Dirac mass of the order of $m_\nu \sim 
v_L^2/v_R$, where-from we extract $v_R \sim 10^{15}$ GeV. If $v_R$ is so large,
the current experimental bounds on the anomalous couplings of gauge bosons 
and the appearance of $Z^\prime$ \cite{zprime} are far away from what is 
expected from the small neutrino masses. The other possibility
is to assume the existence of additional sterile neutrino mass $m_S$, which 
activates the see-saw mechanism, but in the model, where the Dirac mass can be 
essentially reduced from $200$ GeV by several orders of magnitude due to the
suppression $v_L/v_R$.

\section{Hierarchy of scales and GUT}

The arrangement of vacuum expectation values for the spontaneous
breaking of the local gauge symmetries can be reasonably related to 
the following qualitative peculiarities, belonging to the corresponding 
invariant actions. So, we observe:
\begin{itemize}
\item 
The abelian $U(1)$-field, coupled in the vector-like way to the fermions of
both chiralities, does not appear in the spontaneously breaking phase.
\item 
The non-abelian $SU(3)$ field, possessing the asymptotic freedom, is coupled,
again, in the vector-like way to the fermions, and it does not acquire
the spontaneous breaking of symmetry, too. However, the back-wise face of
asymptotic freedom is the confinement.
\item 
The non-abelian $SU(2)$-field, coupled to the chiral fermions, exposes the 
spontaneous breaking.
\end{itemize}
The presence of asymptotic freedom for the latter symmetry depends on the
set of matter fields. In what follows, we exploit the situation, when 
$SU(2)$ is asymptotically free.

Let us offer the following picture. The structure of vector-like gauge
symmetries, i.e. the form of effective potential, preserves them from 
the developing of non-trivial vacuum correlations, determining the spontaneous 
breaking of invariances. The non-abelian theory with chiral fermions
does possess the effective action, where the vacuum correlations appear, if 
the coupling constant is greater than a critical value $\hat\alpha$. 
Then in GUT \cite{gut} with $\alpha_{\rm GUT}< \hat\alpha$, 
the chiral $SU(2)$ will 
develop the symmetry breaking {\sc vev} at a low scale $M_2$,
where its coupling $\alpha_2(M_2)$ will reach the critical value. That can be
the reason for the very different values of $M_{\rm GUT}$ and $M_2$. Note, that
in this approach we know the value of critical constant $\hat\alpha$, since
it is well measured in the weak interactions, so that $\hat\alpha\approx
1/30$.

As for the neutrino mass generation described above, we are ready to conclude,
that the difference between the renormalization group properties, determining 
the coupling running, for the left- and right-handed symmetries will result in
the hierarchy of scales for their characteristic {\sc vev}s. 

To be more concrete, consider the gauge symmetry for the right-handed
fermions, embedded to the following $SU(2)_R$-like doublets:
\begin{equation}
  \label{eq:10}
  \left(e \atop d\right)_R\;\;\;\; \left(\nu \atop u\right)_R
\end{equation}
The essential difference from the usual $SU(2)$ is that the charged
vector bosons, possessing the fractional electric charge $\pm \frac{2}{3}$,
are the color triplet $3_c$ and anti-triplet $\bar 3_c$, appropriately.
The corresponding generalized derivative, acting on the fermions, 
has the form
\begin{equation}
  \label{eq:11}
  iD^\mu = i\partial^\mu- \frac{g_{R}}{2} (\tau_3 Z_{R}^\mu+
 \sqrt{2}\tau_+ W^{i\mu}_{-2/3}+\sqrt{2}\tau_- W^{\mu}_{i\;+2/3}),
\end{equation}
where $\tau_{3,\pm}$ are the Pauli matrices, and the superscript $i$ runs
over the color anti-triplet and the subscript $i$ does the color triplet.

It is quite clear, that, including the quark colors, the "doublets" in
(\ref{eq:10}) could represent the $SU(4)_R$ fundamental multiplet, if 
the couplings of $g_R$ for the right-handed fermions and $g_3$ in $SU(3)$
would be equal each to other. If we "switch off" the color indexes from the
group transformations, the corresponding invariance $G_R$ is
$SU(4)/SU(3)$ on the right-handed fermions. 

As we have mentioned, the $G_R$ symmetry is very similar to the famous $SU(2)$.
The straightforward consideration leads to that we can reproduce the one-loop 
calculations for the running of $g_R$ from the
corresponding evaluation for $SU(2)$, if we substitute for 
$C_F=\frac{(N^2-1)}{2N}=\frac{3}{4}$ at $N=2$ by $\tilde C_F=\frac{1+2N_c}{4}
=\frac{7}{4}$, where $N_c=3$ is the number of colors, and for $C_A=N=2$ by
$\tilde C_A= 2 N_c= 6$.

The running of coupling constants in the $SU(N)$ field theory is given by
the expression
\begin{equation}
  \label{eq:12}
  \frac{1}{\alpha_N(M)} = \frac{1}{\alpha_N(M_0)}+
\frac{b_N}{2\pi}\ln\frac{M}{M_0},
\end{equation}
where $b_N$ depends on the set of fields. So,
\begin{equation}
  \label{eq:13}
  b_N = \frac{11}{3} N -\frac{1}{3} n_f -\frac{1}{6} n_s,
\end{equation}
where $n_f$ is the number of chiral fermions, $n_s$ is the number of
fundamental scalar multiplets. Therefore, for the $b$-coefficients
of $G_R\otimes SU(2)_L\otimes SU(3)$ we get
\begin{eqnarray}
  \label{eq:14}
  b_{R} & = & 22 - \frac{2}{3} n_g - \frac{1}{6} n_{s(R)},\nonumber\\
  b_{L} & = & \frac{22}{3} - \frac{4}{3} n_g - \frac{1}{6} n_{s(L)}, \\
  b_{3} & = & 11 - \frac{4}{3} n_g - \frac{1}{6} n_{s(c)},
\end{eqnarray}
where $n_g$ is the number of fermion generations,
$n_{s(R,L,c)}$ are the numbers of corresponding scalars.

Next, the $U(1)$ coupling constant, normalized as $\alpha_1= \frac{5}{3}
\alpha_{Y}$, where $Y$ is the weak hyper-charge, has the coefficient $b_1$
equal to
\begin{equation}
  \label{eq:15}
  b_1 = -\frac{4}{3}n_g-\frac{2}{15} n_{s(Y)},
\end{equation}
where we have taken into account the hyper-charge of additional weak 
doublet due to the extension of standard higgs by $G_R$:
\begin{equation}
  \left(h_+\atop h_0\right)\to \left(
    \begin{array}{cc}
   h_+ & h_{+1/3} \\
   h_0 & h_{-2/3}
    \end{array}\right),
\end{equation}
where the fractionally charged higgses are the color anti-triplets with the
hyper-charge $Y=-1/3$. So, $n_{s(Y)}$ denotes the number of standard higgses.
\setlength{\unitlength}{1mm}

\begin{figure}[th]
\begin{center}
\begin{picture}(130,80)
\put(0,0){\epsfxsize=12cm
\epsfbox{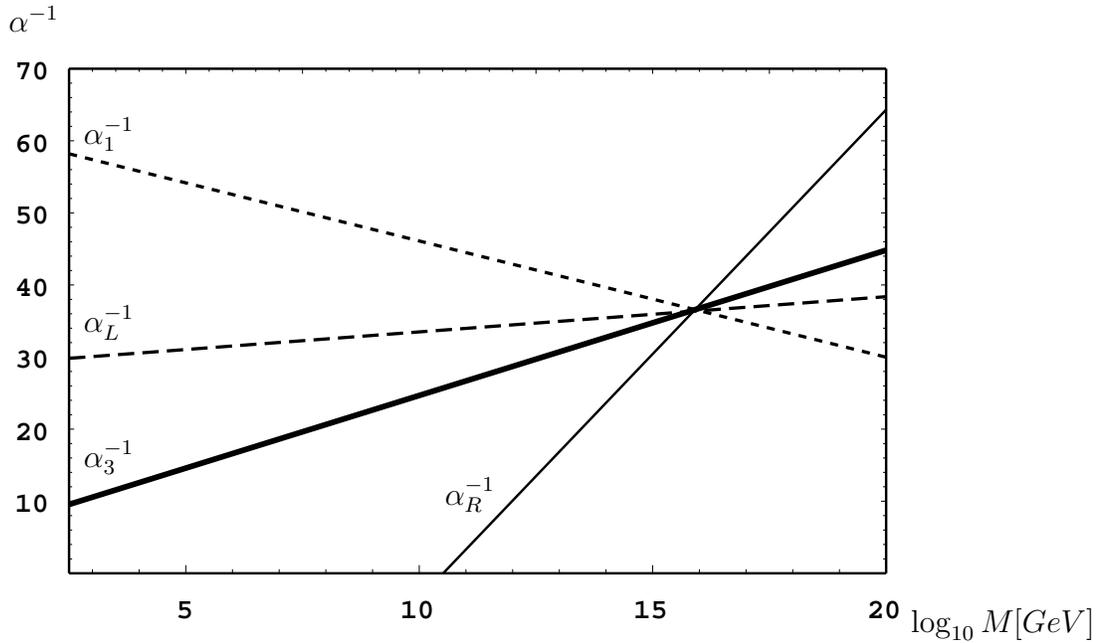}}
\put(0,80){$\alpha^{-1}$}
\put(120,0){$\log_{10}M[GeV]$}
\put(10,65){$\alpha^{-1}_1$}
\put(10,40){$\alpha^{-1}_L$}
\put(10,22){$\alpha^{-1}_3$}
\put(58,17){$\alpha^{-1}_R$}
\end{picture}
\end{center}
\caption{The unification of couplings in GUT with no SUSY.}
\label{gut}
\end{figure}

Let us discuss the scalar field set, suitable for the problem. First, the 
number of left-symmetric higgses includes $1+3$ doublets. In realistic models
for the generation replication and their mixing, the number of standard
higgses usually repeats the number of generations \cite{ng}. So, we put
\begin{equation}
  \label{eq:16}
   n_{s(L)} = 4 n_g,\;\;\;
   n_{s(Y)}= n_g.\;\;\;
\end{equation}
Second, the right-symmetric scalars are those of extensions for
the standard higgs and that of $SU(2)_L$ singlet to separately
break the $G_R$-symmetry. We put
\begin{equation}
  \label{eq:17}
  n_{s(R)} = 3 n_g, \;\;\; n_{s(c)} = 3 n_g.
\end{equation}
Further, we can look at the evolution of couplings to large scales as
it is shown in Fig.\ref{gut}, and draw the conclusion on the plausible 
unification of symmetries at $M_{\rm GUT}\sim 8. \cdot 10^{15}$ GeV. We present
this picture for the illustration of other feature: the $\alpha_R$ coupling
reaches the critical region, $\alpha \simeq \hat \alpha = 1/30$, 
at the scales, which are only one or two orders of magnitude less 
than the GUT energy because of the appropriate properties in the
renormalization group. Of course, the numerical estimate qualitatively 
depends on $\hat \alpha$, which can vary over the structure of 
right-handed symmetry. Another note concerns the extra-higgses, which,
according to the evolution performed, are much lighter than $M_{\rm GUT}$.

As for the model under discussion, we could add only that, obviously,
there is the right-handed symmetry at $M_{\rm GUT}$: $SU(4)_R$ with 
the violation in the way $SU(4)_R\otimes SU(3)_L\to G_R\otimes
SU(3)$.

To complete, we have to emphasize that the correlators in 
(\ref{eq:7})-(\ref{eq:8}) belong to the $(\frac{1}{2},\frac{1}{2})$ 
representation over $G_R\otimes SU(2)_L$, and, hence, contribute, 
a little bit, to {\sc vev} of the extended higgs, and not to 
the $G_R$-doublet, developing the $v_R$ scale itself.

\section{Conclusion}

We have shown how the model of mass generation can be constructed on the
basis of higgs mechanism, wherein the scalar isodoublet field is 
related to the vacuum correlations and fermion charges, so that
\begin{itemize}
\item 
the neutrino is massless in the Standard Model because of its zero electric
charge,
\item 
the smallness of neutrino masses can be caused by the hierarchy of 
the correlation scales for the spontaneous breaking of 
the standard local symmetry and the right-handed
extension, $v_L \ll v_R$, which leads to the Dirac kind of mass,
\item 
the unification of coupling constants makes $v_R$ to be only one or 
two orders of magnitude less than $M_{\rm GUT}$,
\item 
we could reduce the scale of sterile Majorana mass by involving both the
see-saw mechanism and suppressed Dirac terms due to the above approach.
\end{itemize}

Finally, the author would like to express the gratitude to Prof. A.Wagner and
members of DESY Theory Group for their kind hospitality during my visit 
to DESY, where this paper was written, as well as to Prof. A.K.Likhoded 
for discussions.

This work is in part supported by the Russian Foundation for Basic Research.

\end{document}